\documentclass[sigconf, 10pt, nonacm]{acmart}
\setcopyright{none}
\renewcommand\footnotetextcopyrightpermission[1]{}

\usepackage{times}  
\usepackage{hyperref}
\usepackage{latexsym}
\usepackage[export]{adjustbox}
\usepackage{hyperref}
\usepackage{multirow}
\usepackage{tikz}
\usepackage{array}
\usepackage{xspace}
\usepackage{multirow}
\usepackage{comment}
\usepackage{url}
\usepackage{xurl}
\usepackage{listings}
\usepackage[normalem]{ulem}
\usepackage{mdwlist}
\usepackage{algorithm}
\usepackage{booktabs}
\usepackage[noend]{algpseudocode}
\usepackage{centernot}
\usepackage{fancyvrb}
\usepackage{alltt}
\usepackage{subcaption}
\usepackage[ampersand]{easylist}
\usepackage{fancybox}
\usepackage{graphicx}
\usepackage{tcolorbox}
\usepackage{balance}
\usepackage{xcolor}
\usepackage{verbatim}
\usepackage[framemethod=tikz]{mdframed}

\hypersetup{pdfstartview=FitH,pdfpagelayout=SinglePage}

\definecolor{red}{rgb}{0.6,0,0} 
\definecolor{blue}{rgb}{0,0,0.6}
\definecolor{green}{rgb}{0,0.8,0}
\definecolor{cyan}{rgb}{0.0,0.6,0.6}
\definecolor{lightgray}{gray}{0.98}
\definecolor{lightblue}{rgb}{0.13, 0.67, 0.8}
\definecolor{lightorange}{RGB}{255,247,230}
\definecolor{codegreen}{rgb}{0,0.6,0}
\definecolor{codegray}{rgb}{0.5,0.5,0.5}
\definecolor{codepurple}{rgb}{0.58,0,0.82}
\definecolor{keywordcolor}{RGB}{94,20,64}
\definecolor{bluekeywords}{rgb}{0,0,1}
\definecolor{greencomments}{rgb}{0,0.5,0}
\definecolor{redstrings}{rgb}{0.64,0.08,0.08}
\definecolor{xmlcomments}{rgb}{0.5,0.5,0.5}
\definecolor{types}{rgb}{0.17,0.57,0.68}
\definecolor{KWColor}{RGB}{0,0,255}
\definecolor{AnnotationColor}{RGB}{0,137,180}
\definecolor{BlackColor}{RGB}{0,0,0}
\definecolor{CommentColor}{rgb}{0.12,0.38,0.18}
\definecolor{StringColor}{rgb}{0.06,0.10,0.98}
\definecolor{darkred}{rgb}{0.65,0,0}
\definecolor{lightgrey}{rgb}{0.8,0.8,0.8}
\definecolor{marmalade}{RGB}{193,101,18}
\definecolor{peach}{RGB}{250,217,193}
\definecolor{lime}{RGB}{220,237,193}
\definecolor{codeblue}{rgb}{0,0,1}
\definecolor{codegreen}{rgb}{0,0.6,0}
\definecolor{codegray}{rgb}{0.5,0.5,0.5}
\definecolor{codepurple}{rgb}{0.58,0,0.82}
\definecolor{backcolour}{rgb}{0.95,0.95,0.92}
\definecolor{nocolor}{rgb}{1,1,1}

\lstdefinestyle{Terraform}{
  showspaces=false,
  showtabs=false,
  tabsize=2,
  columns=flexible,
  keepspaces=true,
  language={Java},
  numbers=left,
  xleftmargin=0pt,
  basicstyle=\ttfamily\footnotesize,
  commentstyle=\color{CommentColor}\ttfamily\footnotesize,
  stringstyle=\color{codeblue},
  escapeinside={/*@}{@*/},
  numberstyle=\scriptsize\color{gray},
  showstringspaces=false,
  upquote=true,
  xleftmargin=1.2em,
  framexleftmargin=1.5em,
  keywords={ resource }, 
  keywords=[2]{  },
  keywords=[3]{ var, id, network_interface, virtual_machine, nic1, region },
  keywordstyle=\color{BlackColor}\bfseries,
  keywordstyle=[2]\color{codeblue},
  keywordstyle=[3]\color{red},
  moredelim=[il][\color{darkgray}]{$$},
}

\mdfdefinestyle{background}{backgroundcolor=lightorange,innerrightmargin=0cm,innertopmargin=-0.1cm,innerbottommargin=-0.10cm,leftmargin=+0cm, roundcorner=2pt}

\usepackage[textsize=tiny,textwidth=0.6in]{todonotes}
\newcommand{\allnotes}[1]{}
\renewcommand{\allnotes}[1]{#1} 

\newcommand{\yiming}[1]{\allnotes{\todo[color=pink!50]{Yiming: #1}}}
\newcommand{\yh}[1]{\allnotes{\todo[color=cyan!50]{Yibo: #1}}}
\newcommand{\zy}[1]{\allnotes{\todo[color=orange!50]{Zhenning: #1}}}
\newcommand{\fan}[1] {\allnotes{\todo[color=red!30]{Fan: #1}}}

\newcolumntype{C}[1]{>{\centering\arraybackslash}m{#1}}
\newcolumntype{R}[1]{>{\raggedleft\arraybackslash}m{#1}}
\newcolumntype{L}[1]{>{\raggedright\arraybackslash}m{#1}}

\newcommand{\sys}{FlexEMR\xspace}
\newcommand{\emr}{EMR\xspace}

\acmConference[Submitted for review to HotNets]{}
\acmYear{2024}
\copyrightyear{}

\begin{document}


\title{A Disaggregation Approach to Embedding Recommendation Systems}


\author{\fontsize{12}{12}\selectfont Yibo Huang, Zhenning Yang, Jiarong Xing$^\dag$, Yi Dai$^\star$, Yiming Qiu \\ Dingming Wu$^\ast$, Fan Lai$^\diamond$, Ang Chen \vspace{3mm} }

\affiliation{
\institution{\large University of Michigan \hspace{2mm} $^\dag$Rice University \hspace{2mm} $^\star$Fudan University \\ \hspace{2mm}  $^\ast$Unaffiliated \hspace{2mm} $^\diamond$University of Illinois Urbana-Champaign}
    \country{}
}



\begin{abstract}

Efficiently serving embedding-based recommendation (EMR) models remains a significant challenge due to their increasingly large memory requirements. 
Today's practice splits the model across many monolithic servers, where a mix of GPUs, CPUs, and DRAM is provisioned in fixed proportions. This approach leads to suboptimal resource utilization and increased costs. Disaggregating embedding operations from neural network inference is a promising solution but raises novel networking challenges.
In this paper, we discuss the design of \sys for optimized EMR disaggregation. \sys proposes two sets of techniques to tackle the networking challenges: Leveraging the temporal and spatial locality of embedding lookups to reduce data movement over the network and designing an optimized multi-threaded RDMA engine for concurrent lookup subrequests. We outline the design space for each technique and present initial results from our early prototype.

\end{abstract}

\settopmatter{printfolios=true}
\maketitle
\pagestyle{plain}

\section{Introduction}

Embedding-based Recommendation (EMR) models, widely used in e-commerce, search engines, and short video services, dominate AI inference cycles in production datacenters, such as those at Meta \cite{gupta2020architectural, fan2023ada}.
They process user queries using both continuous and categorical features, transforming categorical features into dense vectors via embedding table lookups, and finally, combining them with the continuous features for neural network (NN) scoring.

Serving EMR models at scale leads to pressing memory requirements~\cite{mudigere2022software} as embedding tables can grow to terabytes in size, accounting for over 99\% of model parameters~\cite{desaitrade}.
In practice, we need to partition an EMR model and distribute it across multiple monolithic servers with a mix of GPUs, CPUs, and DRAM~\cite{wei2022gpu, zha2022autoshard, naumov2019deep, fan2023ada}.
On a specific server, 
existing advances propose to decouple the embedding lookup from the NN computation and use DRAM for embedding store~\cite{lui2021understanding}. 
Recent work~\cite{wei2022gpu} further enhances this approach by employing an embedding cache on GPUs to optimize lookup performance.
However, this monolithic approach has limitations in scalability and total
cost of ownership (TCO) in practice.
Recommendation workloads need a mix of resources---memory for embedding store and GPUs for NN computation, and this mixture varies across models and evolves over time. Monolithic servers that provision resources in a fixed portion is hard to achieve both performance and cost efficiency. Recent  studies show that it can lead to idle resources and wasted costs of up to 23.1\%~\cite{DisaggRec}.

A promising approach to achieve performant and cost-efficient large \emr model serving is to disaggregate embedding storage and NN computation into independent servers. 
Specifically, using CPU-based servers to store embedding tables in memory while utilizing GPU nodes for NN computations. These components are interconnected via high-speed networks, such as remote direct memory access (RDMA) \cite{bai2023empowering, shan2018legoos, guo2016rdma} and PCIe interconnect \cite{huang2022ultra}. 
This decouples the memory and GPU resources and allows them to scale independently, improving the total resource efficiency and reducing the TCO. Disaggregation also increases system robustness, as failures are isolated to individual components.

However, disaggregating \emr model serving raises novel networking challenges.
First, remote embedding lookup involves extensive data transmission over the network.
For example, an 8-byte categorical feature index could generate a returned embedding vector with hundreds or even thousands of float values \cite{dlrm-dataset, zha2022dreamshard}. 
Worse still, each lookup needs to query multiple such indices, and each batch contains up to thousands of lookups \cite{naumov2019deep, jain2023optimizing}.
This can be efficiently handled by local GPU memory with high memory bandwidth in a monolithic design. However, decoupling embedding storage and computation shifts this pressure to the network, with a much smaller bandwidth, potentially causing network bottlenecks. 
On the other hand, intensive data transmission imposes stringent performance requirements on the network layer. Unfortunately, today's RDMA systems \cite{gao2021cloud, xue2019fast, huang2019bor, huang2019rdma, liu2023remote} are not designed for \emr disaggregation. For instance, the single-thread RDMA I/O models that are commonly used in regular applications \cite{nsdi19-erpc, nsdi2014farm, } will suffer from high software queuing latency for EMR serving. 
The recent design on disaggregated EMR systems~\cite{DisaggRec} mainly focuses on resource provisioning but overlooks the above networking challenges.

In this paper, we design an optimized disaggregated \emr system called \sys. \sys optimizes the disaggregation by proposing two classes of techniques to tackle the challenges discussed above.
The first set of techniques explores the \textit{temporal} and \textit{spatial} locality of embedding lookup. 
While existing works~\cite{wei2022gpu} implement embedding caches on GPUs to leverage temporal locality,  
we observe that such caches could compete with NN computation for limited GPU memory, and propose mechanisms to dynamically adjust caching strategy to avoid contention.
Given that multiple lookup subrequests could point to the same embedding server, we further investigate the benefit of spatial locality.
we design a hierarchical embedding pooling strategy that partially offloads pooling operations into the embedding servers, utilizing their available CPU cycles. This reduces embedding movement in the network and mitigate pressure on the rankers, while putting CPU resources to good use.

The second set of techniques aims at optimizing the RDMA I/O engines used for remote embedding lookups. First, we explore the design of a contention-free multi-threaded RDMA service, allowing concurrent RDMA threads to post lookup requests to different embedding servers in parallel. This approach significantly reduces queuing latency compared to commonly used single-threaded RDMA solutions. Additionally, we handle skewed access patterns by periodically migrating connections across embedding servers, and deploy credit-based flow control to mitigate response congestions. These optimizations collectively enhance the performance of remote embedding lookups in \sys.

Working together, these two sets of techniques enable an efficient, flexible, and cost effective \emr model serving architecture. We validate the key ideas of \sys using micro-benchmarks and present preliminary results in \S\ref{sec:eva}.

\section{Overview}





In this section, we provide background on EMR model serving, describe the motivation and challenges for disaggregated \emr serving, and present an overview of our solutions.

\subsection{Background: EMR models}

\begin{figure}[!t]
	\centering
	\includegraphics[width=0.99\linewidth]{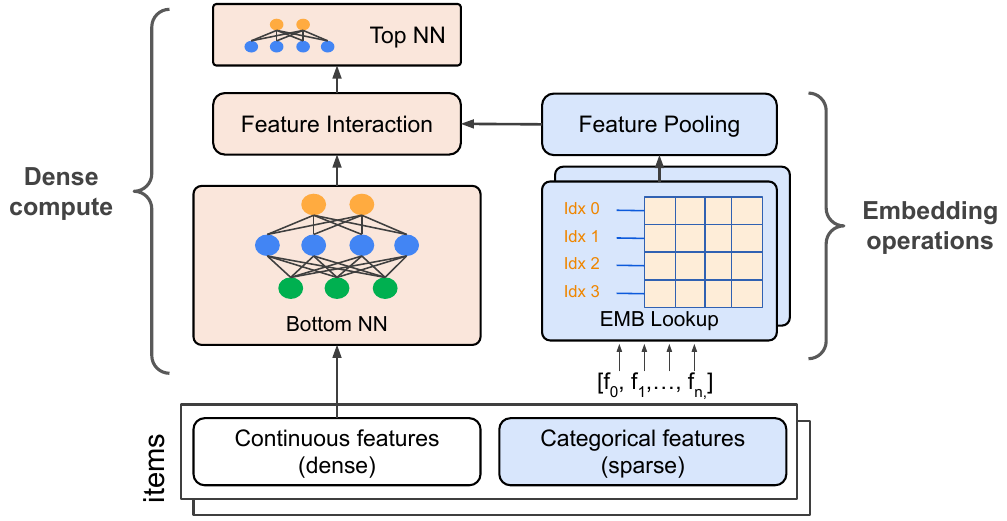}
	\caption{A representative EMR model---Deep Learning Recommendation Model (DLRM).}
	\label{fig:emr-model}
	\vspace{-0.3cm}
\end{figure}

An \emr model handles two types of input features: \textit{categorical} features (\textit{sparse}) representing discrete categories or groups and \textit{continuous} features (\textit{dense}) representing measurements or quantities that are continuous in nature \cite{naumov2019deep}. For example, in a video recommendation system, categorical features could include video IDs, genres, or user IDs, while continuous features could include user age or watch time. 
The categorical features often have very high cardinality, as each feature can consist of millions of instances (e.g., numerous specific IDs for users in feature ``user IDs''). 
\emr models convert these high-dimensional categorical features into dense vector representations via \textit{embedding tables}.


Figure~\ref{fig:emr-model} illustrates the key components and workflow of a representative EMR model. It takes candidate items as input, including both categorical and continuous features from upstream. 
For those accessed instances (e.g., IDs) in a batch, \emr retrieves their associated embeddings (dense vectors), which will be aggregated into a single fixed-size embedding vector through pooling operations such as sum or average. 
Meanwhile, the continuous features are processed by a bottom neural network (\textit{bottom NN}) which is typically a multilayer perceptron (MLP) to generate high-dimensional dense vectors. 
The feature interaction process combines the dense vectors from categorical and continuous input features through operations such as element-wise multiplication or concatenation. 
The combined result is fed into a top neural network (\textit{top NN}) to compute user-item scores for top-k ranking. 
The items with the highest scores are presented to the user.





\subsection{Motivation: Disaggregated EMR serving}


State-of-the-art EMR models consist of hundreds of sparse features, each associated with an embedding table with potentially millions of embedding rows~\cite{isca20-dlrm, asplos23-embedding-column}.
Indeed, production-level \emr models could have TB-level embedding tables (e.g., Meta uses 50TB DLRM model~\cite{mudigere2022software}).

The large-size embeddings have presented significant challenges for EMR serving because they cannot be stored on a single GPU.
Therefore, EMR embeddings are often partitioned and scattered across multiple servers, each server has a combination of GPUs, CPUs, and DRAM.
Considering that EMR workloads require two distinct types of resources—large memory for embedding storage and GPUs for NN computation—researchers propose decoupling them for better flexibility. Specifically, this approach leverages DRAM and CPUs for embedding storage and lookup, while utilizing GPUs on the same servers for NN computation.
As a further improvement, an embedding cache is employed in GPU memory to cache the ``hot'' entries to optimize the lookup performance~\cite{wei2022gpu}.

The embedding-NN decoupling enables more flexible \emr serving, but doing that on monolithic servers has several limitations.
Monolithic servers provision GPU, CPU, and DRAM resources in fixed proportions, but the demands for these resources by EMR workloads can evolve across models and change over time due to varied recommendation workloads.
Scaling up the whole server for the most demanding resource or the peak workload will lead to low resource utilization and waste of costs.
A recent study~\cite{DisaggRec} has found that fixed resource provision on monolithic servers can result in wasted costs of up to 23.1\%. 
Therefore, more resource- and cost-efficient \emr serving are urgently needed.

Building upon the trend of disaggregation in datacenters~\cite{10056149, DxPU}, a promising solution is to fully disaggregate the embedding layer and dense NN compute into network-interconnected CPU embedding servers and GPUs (rankers), respectively. 
The rankers access embeddings stored on embedding servers over high-speed networks, such as RDMA.
This new EMR serving paradigm offers multifold benefits including: (1) Flexible scalability. It allows each component to scale independently, e.g., allocating additional memory to accommodate larger embedding tables. 
(2) Cost efficiency. It allows rankers to multiplex many embeddings streamed from CPU embedding servers, greatly improving resource utilization and reducing the number of rankers needed for deploying \emr models.
(3) Improved robustness. It enhances system robustness because the embedding operations and NN computation failures can be perfectly isolated.

EMR disaggregation is an emerging direction that remains underexplored. The most closely related work, DisaggRec~\cite{DisaggRec}, shares similar disaggregation concepts with us but primarily focuses on resource provisioning and scheduling post-disaggregation. However, EMR disaggregation introduces several networking challenges, as we will discuss next, that have not yet been thoroughly studied.

\begin{figure}[!t]
	\centering
        \includegraphics[width=0.8\linewidth]{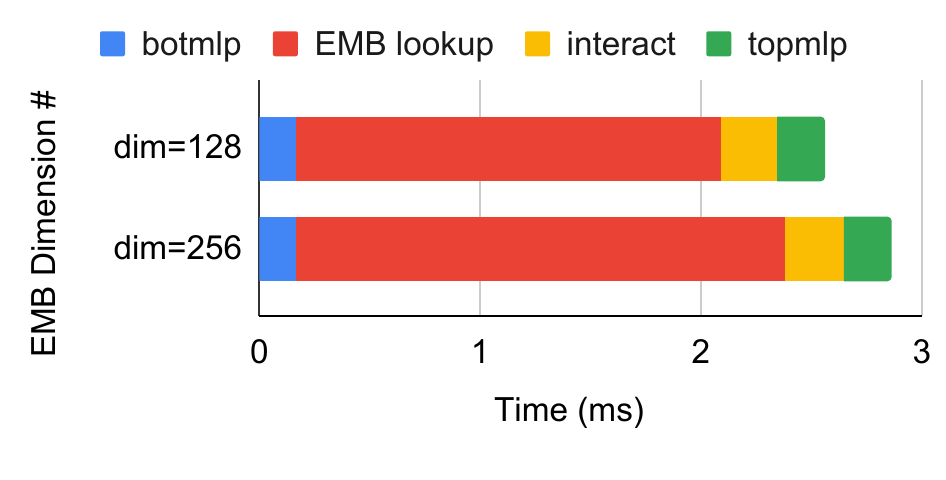}
        \vspace{-0.5cm}
	\caption{Embedding layer dominates EMR serving.}
	\label{fig:emb-dominated}
	\vspace{-0.5cm}
\end{figure}

\subsection{Key research challenges }






Disaggregated EMR serving involves a large volume of data movements over the network. 
Typically, given some categorical feature indices as input, the ranker first fetches all corresponding embedding vectors from remote embedding servers and then performs feature pooling operations. 
As a result, the network bandwidth between embedding servers and rankers becomes a major bottleneck (as shown in Figure~\ref{fig:emb-dominated}), presenting several domain-specific challenges.


\vspace{1mm}
\noindent\textbf{(1) Contention in GPU memory.}
To reduce data movement over the network, existing works~\cite{Kal2021cache, Ibrahim2021cache, ke2020recnmp, Lee2021cache, Wilkening2021cache, xie2022fleche, kurniawan2023evstore, kwon2019tensordimm} attempt to cache frequently accessed embedding entries in GPU memory.
However, we observe that embedding cache is far from a perfect solution. 
Using precious GPU memory for caching could significantly reduce serving throughput, especially when the NN model size and request batch size are large. Essentially, NN inference also requires a large amount of GPU memory, and the existing caching strategy could cause serious resource contention between the two tasks.


\vspace{1mm}
\noindent\textbf{(2) Large-scale fan-out pattern.} 
Remote embedding lookup generates large-scale fan-out subrequests. 
For example, an 8-byte categorical feature index could generate a returned embedding vector with hundreds or even thousands of bytes in dimension size. Moreover, each lookup needs to query multiple such indices, and each batch contains up to thousands of lookups. 
Unlike local memory, the network bandwidth is significantly lower. 
Hence, issuing hundreds or thousands of concurrent batched embedding lookups can lead to severe network contention and degraded performance. 

\vspace{1mm}
\noindent\textbf{(3) RDMA engine efficiency.}
RDMA is commonly used for remote data access. Most existing RDMA applications employ single-threaded RDMA I/O models, which send out RDMA read requests to different target machines using one thread.
This leads to extended queuing latency in our scenario. We need to design a more efficient RDMA I/O engine capable of handling concurrent embedding lookup requests and results, while effectively re-balancing skewed workload patterns across distributed embedding servers.

\begin{figure}[!t]
	\centering
	\includegraphics[width=0.95\linewidth]{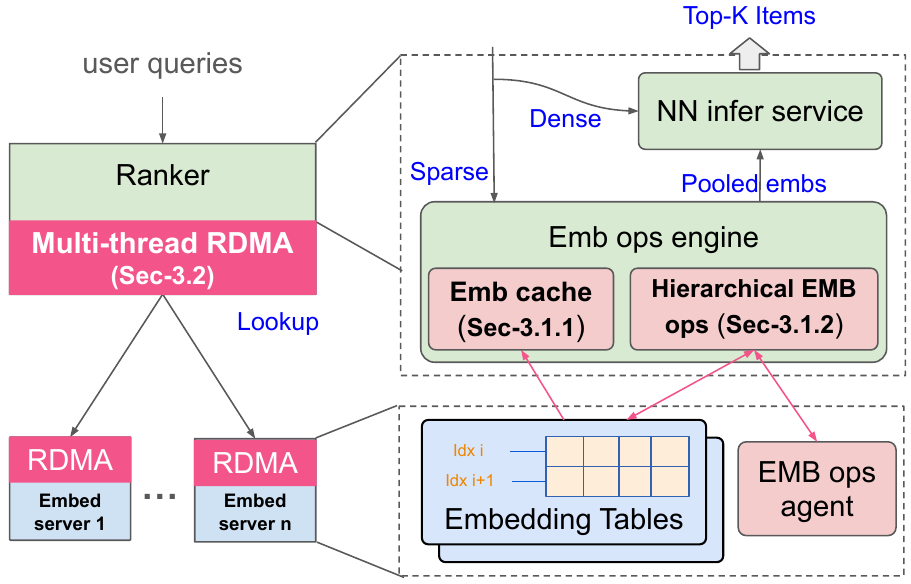}
	\caption{\sys architecture overview.}
	\label{fig:flexemr-arch}
	\vspace{-0.5cm}
\end{figure}

\subsection{Our solution: \sys}




In this paper, we propose \sys---an \emr serving system that aims at addressing the aforementioned challenges. 
Figure~\ref{fig:flexemr-arch} illustrates the envisioned system architecture. At a high level, the rankers initiate embedding lookups. Each lookup contains multiple subrequests, which firstly go through an adaptive embedding cache on rankers serving as a lookup fast path for reduced latency.
The requests are then sent to embedding servers via a set of optimized RDMA engines. Once the corresponding embedding vectors are found, \sys initiates a hierarchical pooling process to retrieve results without causing network contentions. 


Our design is primarily based on two sets of techniques.
First, we reduce embedding movement over the network by exploiting \textit{temporal} and \textit{spatial} locality across embedding lookups and subrequests. 
Temporal locality means that (i) a non-negligible portion of embeddings (e.g., 10\%$\sim$15\%) are the most frequently accessed in a period (i.e. \textit{hot embeddings} \cite{isca20-dlrm, DeepRecSys2020, Wilkening2021cache}), and (ii) some subrequests often appear together in the same lookup (i.e. \textit{embedding co-occurrence} \cite{haojie2023grace}).
Existing work has leveraged temporal locality to design embedding caches on GPUs, but we argue that the caching design should be dynamically adjusted to avoid GPU memory contention.
Spatial locality means that multiple embedding tables/shards often co-locate in the same embedding server, so many subrequests in a lookup will be sent to the same destinations.
As such, we propose to push-down lightweight pooling operations onto embedding servers. 
This leverages the fact that 
embedding servers also contain CPU resources, but they are under-utilized at runtime~\cite{Jain2023cpu}. 

Second, we improve the networking layer of \sys serving with a fleet of RDMA optimizations. Specifically, we propose to use multi-thread RDMA for embedding lookups. It allows multiple RDMA IO threads to concurrently handle lookups into different embedding servers, thus reducing the queuing latency. We address the RNIC resource contention problem across concurrent RDMA requests using a mapping-aware RDMA IO engine, and envision a load-balanced live migration mechanism to overcome the problem of skewed requests across embedding servers. A new credit-based flow control mechanism is further implemented to avoid head-of-line blocking caused by traffic bursts.

\begin{figure}[!t]
	\centering
	\includegraphics[width=0.99\linewidth]{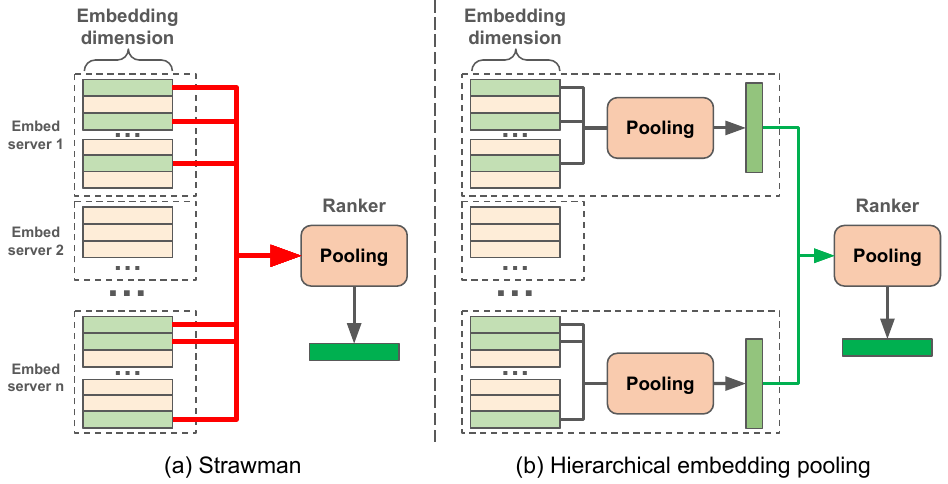}
	\caption{Hierarchical EMB pooling. Pooling computation handled solely by the ranker can cause network contention (\ref{fig:offloading}a). Performing pooling hierarchically, sending only the intermediate results to the ranker can reduce network traffic (\ref{fig:offloading}b). }
	\label{fig:offloading}
	\vspace{-0.53cm}
\end{figure}

\section{\sys design}


In this section, we outline a potential system design and optimizations. 
We first highlight the design of an adaptive caching mechanism and a hierarchical EMB pooling architecture in section~\ref{subsec:locality-disaggr}.
Next, in section~\ref{subsec:multi-thread-rdma}, we discuss how multi-threaded RDMA further optimizes the embedding lookup.


\subsection{Locality-enhanced disaggregation}
\label{subsec:locality-disaggr}

\subsubsection{Adaptive EMB caching}
\label{subsec:hot-emb-cache}

A common practice to reduce embedding lookup latency is to leverage the \textit{temporal locality} across requests and cache ``hot'' lookup or pooling results in GPU memory~\cite{wei2022gpu}. 
However, because the embedding caches share GPU memory with NN computation, an enlarged cache inevitably leads to a smaller batch size for NN computing due to GPU memory contention, thereby degrading overall throughput. In this work, we explore an approach to adaptively adjust the size of cache: when the system is overloaded, \sys reduces cache size automatically to preserve overall throughput; otherwise, it expands the cache to improve latency. 

\begin{figure}[!t]
	\centering
        \includegraphics[width=0.99\linewidth]{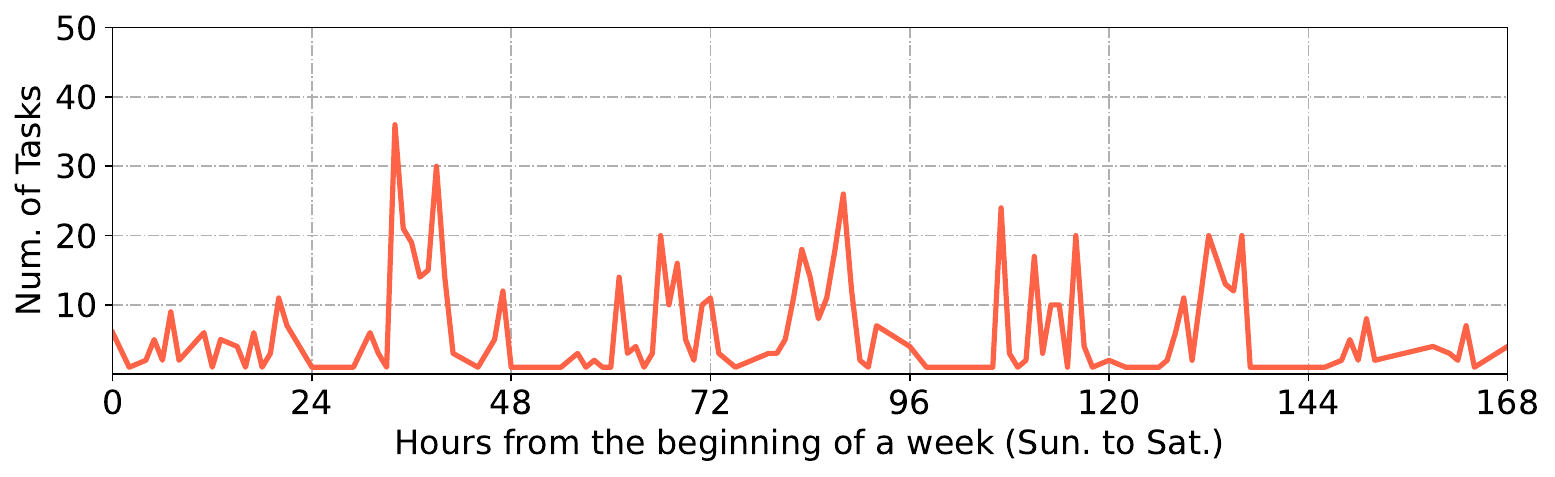}
        \vspace{-0.3cm}
	\caption{Distribution of inference workloads in Alibaba PAI platform over one week.}
	\label{fig:ml-temporal-dynamics}
	\vspace{-0.3cm}
\end{figure}

\vspace{1mm}
\noindent\textbf{Tracing temporal dynamics.}
The first step towards an adaptive caching strategy is to capture the workload temporal dynamics (Figure~\ref{fig:ml-temporal-dynamics}).
In reality, the ranker often uses a task queue to receive batches of requests from upstream, then feeds them into downstream \emr models. 
\sys could monitor the size of these batches, then apply a sliding window algorithm to determine whether the system is under high load.


\vspace{1mm}
\noindent\textbf{Adjusting cache size.}
Once a decision is made to enlarge or shrink cache size, we need to consider how to 
enforce these actions accordingly. This involves two sub-tasks: Firstly, we need to determine the updated cache size. Our observation here is that, given the incoming batch size and the \emr model architecture, it is possible to build a model to estimate the memory size required by NN computation. The ideal cache size is the difference between GPU memory capacity and the parts reserved for NN.  
The second task is to
swap embeddings into or out of GPU memory
For the \textit{swap in} action, \sys could initiate RDMA reads from the ranker to asynchronously fetch the hot embeddings from embedding servers in a transparent manner. 
For the \textit{swap out} action, \sys should remove part of embedding cache lines based on a LRU algorithm, and free up the corresponding GPU memory.

\subsubsection{Hierarchical EMB pooling}
\label{subsec:hierarchical-emb-ops}

Apart from temporal locality, \textit{spatial locality} is also prevalent in \emr serving systems.
In a disaggregated architecture, embedding tables are placed onto a set of remote embedding servers. 
Given an embedding lookup request from the ranker, a typical workflow is shown in Figure~\ref{fig:offloading}(a):
First, the ranker sends sub-requests to remote embedding servers and asks them to return corresponding embedding vectors. 
The ranker then aggregates these results through \textit{pooling} operations. 
This communication leads to extensive embedding movement over the network and increased latency. 

\vspace{1mm}
\noindent\textbf{Hierarchical pooling leveraging spatial locality.}
We seek to reduce the embedding movements between the ranker and embedding servers for higher throughput under bounded latency.
Our finding here is that the CPUs in embedding servers could be utilized to perform \textit{partial pooling} operations. If an embedding server contains multiple required vectors (i.e. spatial locality), then it could aggregate them first before sending to the ranker.
Motivated by this finding, we envision a hierarchical pooling architecture, as shown in \ref{fig:offloading}(b).  
For each embedding lookup, \sys first invokes embedding server CPUs to perform partial pooling whenever possible, then asks the ranker to retrieve their outputs and perform \textit{global pooling} to obtain the final results.
Unlike existing works \cite{zhang2022optimizing},
\sys is the first to explore parallel operator push-downs (i.e., pushing pooling operations down to embedding servers). 
This design could potentially generalize to other workloads with large-scale fan-out patterns.

\vspace{1mm}
\noindent\textbf{Routing table for identifying co-located embeddings.}
An important question here is how to identify the embedding \textit{spatial locality} among embedding servers---i.e., given as input a set of sparse feature indices, we need to identify which indices are co-located at where. 
A na\"ive solution is to maintain a routing table storing the 
\textit{<feature indice, dest embedding server>} mapping pairs. 
It then queries all corresponding embedding servers and aggregates the sparse feature indices into multiple groups depending on their embedding servers. 
However, this leads to huge memory footprints due to numerous sparse feature spaces.
We observe a large embedding table is often partitioned into multiple shards in a row-wise manner, and each shard corresponds to an embedding range containing \textit{start} and \textit{end} indices.
Based on this, we envision a range-based routing table in ranker where we store \textit{<(start index, end index), dest embedding server>} pairs for each embedding shard.
For a list of sparse indices, \sys only needs to use the range to which each indice belongs to get the target embedding server efficiently.

\subsection{EMB lookup with Multi-threaded RDMA}
\label{subsec:multi-thread-rdma}

\begin{figure}[!t]
	\centering
	\includegraphics[width=0.98\linewidth]{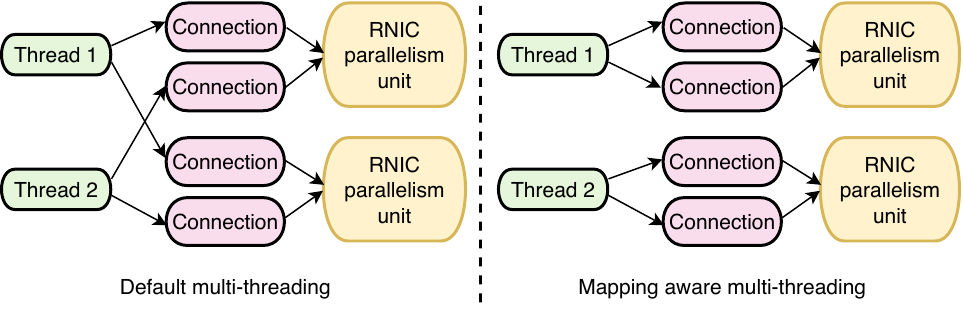}
	\caption{\sys multi-threaded embedding lookup. Mapping awareness eliminates inter-thread contentions.}
	\label{fig:multi-thread-mapping}
	\vspace{-0.5cm}
\end{figure}



We next discuss how to optimize the RDMA I/O engine for remote embedding lookup.
Given a batch of embedding lookup requests, each containing a large amount of fan-out subrequests, rankers in \sys require an efficient RDMA I/O engine to forward subrequests to remote embedding servers. Since the completion time of an embedding lookup is dominated by the slowest subrequest, the overall pipeline is very sensitive to tail latency. The single-threaded RDMA IO model used for most existing RDMA applications \cite{nsdi2014farm} becomes a major bottleneck, since it incurs high queuing latency overhead between the embedding and transport layers.

A promising solution is to use \textit{multi-threaded RDMA}, where RDMA connections to embedding servers are assigned to multiple I/O threads, and embedding subrequests are distributed to these connections according to corresponding destinations.
However, na\"ively using multi-threaded RDMA introduces non-negligible contentions due to limited RNIC parallelism resources (e.g., user access regions \cite{MPI-threads}):
Figure~\ref{fig:rdma-engine} (left) shows that it can lead to up to 62\% throughput drop in our microbenchmark.


\vspace{1mm}
\noindent\textbf{Contention-free multi-threaded embedding lookup.}
To understand the root cause of contentions under concurrent lookup subrequests, we delve deep into the architecture of multi-threaded RDMA. As shown in Figure~\ref{fig:multi-thread-mapping}, we find that each RDMA engine contains a dedicated I/O thread, and each thread encompasses multiple RDMA connections.
The RNIC parallelism units are allocated to each newly created connection in a round-robin manner, resulting in a one-to-many mapping between RDMA parallelism units and connections. 
However, the I/O threads for remote embedding lookup are not aware of such mappings, 
thereby enforcing multiple RDMA connections belonging to different I/O threads to access the same parallelism unit simultaneously. 
To coordinate different I/O threads, each parallelism unit must implement a complex locking mechanism, which could introduce significant performance overhead.

To solve this problem, we envision a mapping-aware multi-threaded RDMA engine, capable of transparently generating one-to-one mapping between I/O threads and RNIC parallelism units, as shown in Figure~\ref{fig:multi-thread-mapping} (right).
The key-enabling technique is the  \textit{resource domain} feature provided by RDMA \cite{resource-domain-rdma, resource-domain-verbs}, which exposes the mapping between connections and RNIC parallelism units to the application layer.
As such, \sys could ensure that all connections assigned to the same parallelism unit are allocated to the same RDMA engine, thus preventing contention from concurrent threads.
Essentially, in the cluster initialization stage, \sys firstly creates RDMA connections between embedding servers and rankers, then identifies their resource domains. Since there is a static mapping between the resource domain and parallelism unit, \sys could subsequently aggregate connections into different RDMA engines according to the resource domain, so that each RDMA engine points to a dedicated parallelism unit.

\vspace{1mm}
\noindent\textbf{Live connection migratation among RDMA engines.}
Another common problem in practice is skewed subrequest patterns.
Connections to different embedding servers might experience vastly different utilization rate, which leads to imbalanced loads among RDMA engines. 
Since an RDMA engine can manage multiple connections used for different embedding servers,
a strawman solution is to live-migrate connections in overloaded engines to 
under-utilized engines: Periodically, \sys traces the number of queued subrequests in each connection. When a connection becomes overloaded, \sys selects the least loaded RDMA engine and initiate the migration process.
However, this workflow brings back the RDMA multi-thread contention problem, because the migrated connection still points to the old parallelism unit.
\sys aims at a live migration strategy without RDMA contention concerns. The key idea is to \textit{re-associate} the migrated connection with the resource domain used by the new RDMA engine. 
Notably, \sys detaches the connection from the old domain, and then attaches it to the resource domain of the new one.


\vspace{1mm}
\noindent\textbf{Fast credit-based flow control with RDMA QoS.}
\sys pipelines pooling computation and remote embedding lookup to further reduce serving latency.
To do that, \sys introduces a per-connection task queue between remote embedding servers and RDMA engines, 
so that the responses (i.e., embedding vectors) from embedding servers can be pushed into their respective queues asynchronously. 
However, without careful flow control, concurrent subrequests could result in response bursts, which might overflow corresponding task queues and drastically increase tail latency.
A strawman solution is to leverage existing credit-based flow control \cite{ATC22-Ali-ZERO, Weibai-credit-work}---specifically, the ranker controls the size of each task queues via \textit{credits} and proactively piggybacks the credits in lookup requests.
However, since these credit messages share channels with regular lookup messages, the latter could easily introduce head-of-line blocking and delay reception of the former. As a result, embedding servers won't be able to adjust sending rate of responses in time.
To mitigate such head-of-line blocking, we envision a fast credit control channel with higher priority.
By leveraging the hardware feature of connection-level quality of service (QoS) offered by RDMA, \sys can create a dedicated RDMA connection with a higher service level for each \textit{<ranker, embedding server>} pair. At runtime, \sys can use such connections as fast path to transfer credits timely even under high load pressure. 
\section{Preliminary results}
\label{sec:eva}

\begin{figure}[t!]
        \vspace{-0.1cm}
	\centering
        \includegraphics[width=0.8\linewidth]{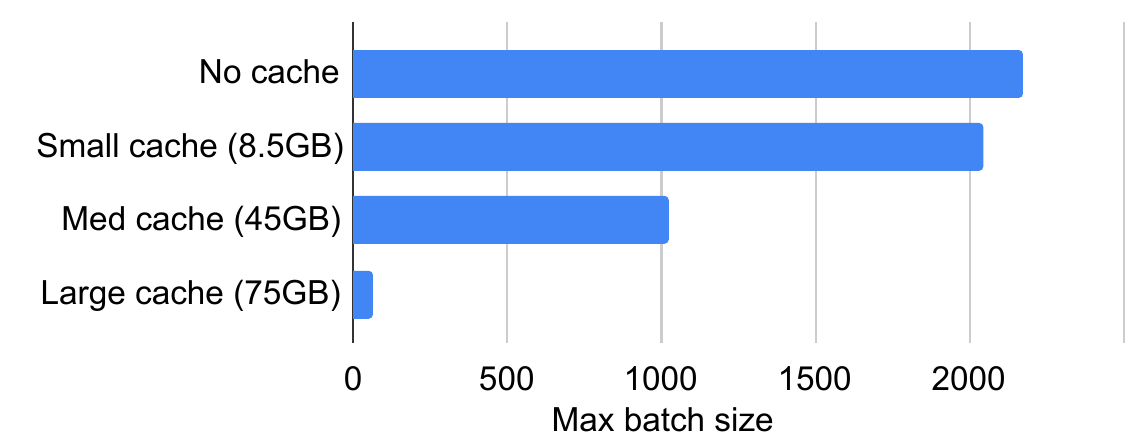}
        \vspace{-0.3cm}
	\caption{GPU caching significantly limits the maximum batch size for inferences due to EMB-NN contentions.}
	\label{fig:emb-cache-batch}
	\vspace{-0.2cm}
\end{figure}

\begin{figure}[t!]
    \begin{minipage}[t]{0.55\linewidth}
        \centering
        \includegraphics[width=1.0\textwidth]{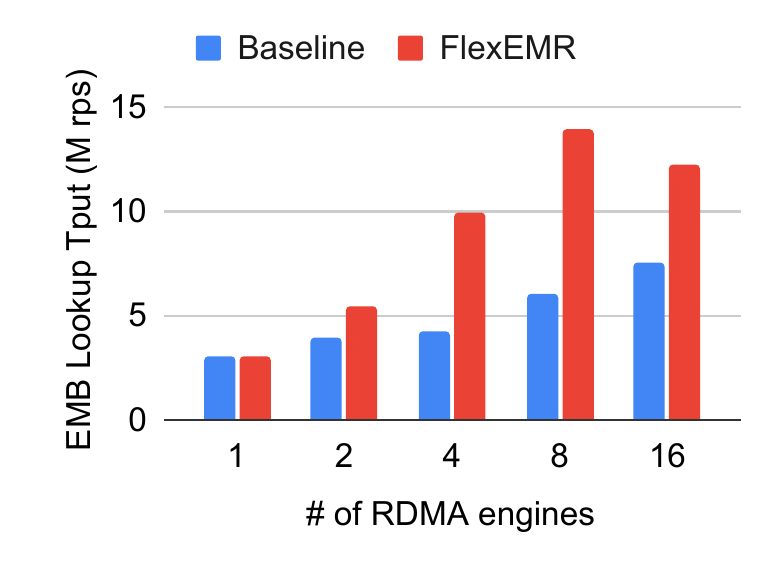}
    \end{minipage}%
    \hspace{0.1cm}
    \begin{minipage}[t]{0.39\linewidth}
        \centering
        \includegraphics[width=1.0\textwidth]{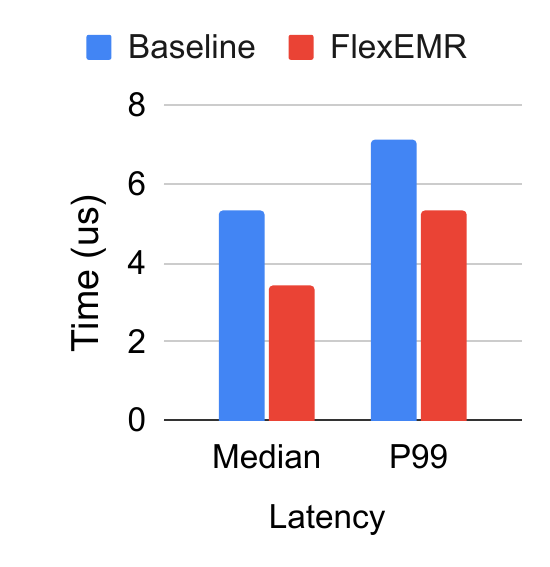}
    \end{minipage}
    \vspace{-0.4cm}
    \caption{Performance comparison between baseline and \sys: Multi-threaded EMB lookup (left) and credit flow control for lookup (right). }
    \vspace{-0.55cm}
    \label{fig:rdma-engine}
\end{figure}

Even though \sys is still work-in-progress, we showcase initial evidence around its major components, including adaptive embedding caching (\S\ref{subsec:hot-emb-cache}) and multi-threaded RDMA embedding lookup (\S\ref{subsec:multi-thread-rdma}). 
We use the popular MLPerf framework \cite{mlperf} and a set of production-scale embedding lookup traces released by Meta \cite{dlrm-dataset} to synthesize inference workloads. 
Our testbed includes two interconnected Intel Xeon servers each equipped with 32 CPU cores, 128GB memory, and a 100Gbps Mellanox RDMA NIC. One of them is equipped with a Nvidia A100 GPU with 80GB of memory. 

\vspace{1mm}
\noindent\textbf{Na\"ive caching leads to GPU contentions.}
To understand the benefit of adaptive EMB caching,
we analyze a pure GPU caching-based solution on a representative RMC2 model~\cite{Jain2023cpu, isca20-dlrm}.
For the GPU caching baseline, we vary the size of EMB caches and observe the changes on supported batch sizes. 
As Figure \ref{fig:emb-cache-batch} shows, as we increase the GPU cache size, the caching-based solution has to settle with smaller batch size due to contention on GPU memory capacity, resulting in decreased inference throughput and wasted GPU compute cycles. 
\sys on the other hand aims to achieve the highest batch size through an 
adaptive embedding caches, 
as proposed in \S\ref{subsec:hot-emb-cache}, mitigating memory contention in most scenarios.



\vspace{1mm}
\noindent\textbf{\sys outperforms na\"ive RDMA-based embedding lookup in efficiency.}
Next, we compare the lookup performance of a na\"ive multi-threaded RDMA baseline against our \sys prototype. 
As Figure \ref{fig:rdma-engine} illustrates, 
with mapping aware multi-threading, \sys achieves higher throughput than baseline by up to 2.3x. 
Moreover, \sys achieves 35\% lower latency on credits transmission, which further reduces possible congestion between rankers and embedding servers. 
This demonstrates the importance of an efficient multi-threaded RDMA engine (\S\ref{subsec:multi-thread-rdma}).



\section{Related Work}




Many existing works treat \emr as generic deep learning models and adopt GPU-centric approaches for their deployment \cite{wei2022gpu, zha2022autoshard, naumov2019deep, fan2023ada}, leading to under-utilized GPU resources.
Recent projects apply a variety of caching mechanism\cite{Kal2021cache, Ibrahim2021cache, ke2020recnmp, Lee2021cache, Wilkening2021cache, xie2022fleche, kurniawan2023evstore, kwon2019tensordimm}  to speed up embedding lookups.
However, these solutions suffer from low cache hit rate in production environments
~\cite{Jain2023cpu}. 
Specialized hardware such as FPGAs has also been explored to enhance recommendation systems \cite{zeng2022faery, Jiang2021fpga, Hsia2023tpu},
but we strive for a generic solution with commodity hardware.
Compression~\cite{MLSYS2021_1c4534ff, MLSYS2022_1eb34d66, 9517710, 10.1145/3394486.3403059} and sharding~\cite{sethi2022recshard, sethi2023flexshard, zha2022autoshard, shi2020compositional, NEURIPS2022_62302a24} are common optimizations to embedding table lookup. 
These works are complementary to ours, as the proposed techniques can be integrated seamlessly into \sys for further improved performance. 
DisaggRec \cite{DisaggRec} proposed a similar disaggregated memory system. However, the resource distribution is fixed and determined through an exhaustive search. This approach introduces overhead and fails to capture serving dynamics.

\section{Conclusion \& Future work}

Embedding-based recommendation (EMR) model serving consumes the majority of AI inference cycles in production datacenters due to its unique embedding-dominated characteristics and stringent service-level objectives.
However, prior serving systems for EMR models struggle to achieve high performance at low cost. 
We propose \sys, a fast and efficient system that disaggregates embedding 
table lookups from NN computation.
\sys uses a set of locality-enhanced optimizations atop a multi-threaded RDMA engine to ensure performance and resource efficiency.
We envision \sys to
improve user experience of Internet-scale recommendation services
while driving down costs for their providers.
Furthermore, we believe this paradigm can benefit other ML workloads, including large language models (LLM) \cite{PagedAttention, orca}, multimodal models \cite{DISTMM}, and mixture-of-expert (MoE) \cite{MoE1, MoE2}, which we will also investigate in future works.

\bibliographystyle{ACM-Reference-Format}
\bibliography{reference}

\end{document}